\documentclass{aa}
\usepackage{graphicx}
\usepackage{times}
\usepackage{natbib}

\begin{document}

\title{The radio spectrum and magnetic field structure of SNR HB3}

\author{W.~B. Shi\inst{1,3} \and
      J.~L. Han\inst{1} \and
      X.~Y. Gao\inst{1} \and
      X.~H. Sun\inst{1} \and
      L.~Xiao\inst{1} \and
P. Reich\inst{2} \and
W. Reich\inst{2}
}
%
%
\institute{National Astronomical Observatories, Chinese Academy of
        Sciences, Jia 20 DaTun Road, ChaoYang District,\\ Beijing 100012,
        China \\ e-mail: {\tt hjl@bao.ac.cn}
 \and
         Max-Planck-Institut f\"{u}r Radioastronomie,
             Auf dem H\"ugel 69, 53121 Bonn, Germany 
 \and
         School of Space Science and Physics, Shandong University
at Weihai, 180 Cultural West Road, Shandong 264209, China}
   \date{Received 29 April 2008 / Accepted 7 June 2008}
%
\abstract
%
{Evidence for a spectral flattening of the supernova remnant (SNR) HB3
(G132.7+1.3) was recently claimed in literature based on previously
published total flux density data, and the flattening was further
interpreted as the discovery of thermal bremsstrahlung emission in the shell
of HB3.}
%
{A spectral flattening has never been observed from any SNR before.
Reliable observations of HB3 at frequencies above 3000~MHz are crucial to
confirm such a spectral behaviour.}
%
{We extracted 4800~MHz total intensity and polarisation data of HB3 from the
Sino-German 6~cm polarisation survey of the Galactic plane made with the
Urumqi 25~m telescope, and analysed the spectrum of HB3, together with
Effelsberg data at 1408~MHz and 2675~MHz.}
%
{We found an overall spectral index of HB3 of $\alpha=-0.61$$\pm$$0.06$
between 1408~MHz and 4800~MHz, similar to the index at lower
frequencies. There is no spectral flattening at high frequencies. We
detected strong polarised emission from HB3 at 4800 MHz. Our 4800~MHz data
show a tangential field orientation in the HB3 shell.}
{} 
%
\maketitle

\section{Introduction}

Radio emission from a supernova remnant (SNR) originates from synchrotron
emission of relativistic electrons gyrating in the enhanced magnetic fields
in its shell or filaments. The observed radio spectrum of a SNR generally
follows a power-law ($S_\nu \propto \nu^{\alpha}$, where $\alpha$ is the
spectral index). A definite case of a spectral break was detected for the
SNR S147 \citep{fr86,xfrh08}.

\begin{figure*}
\centering
\includegraphics[angle=270,width=0.30\textwidth]{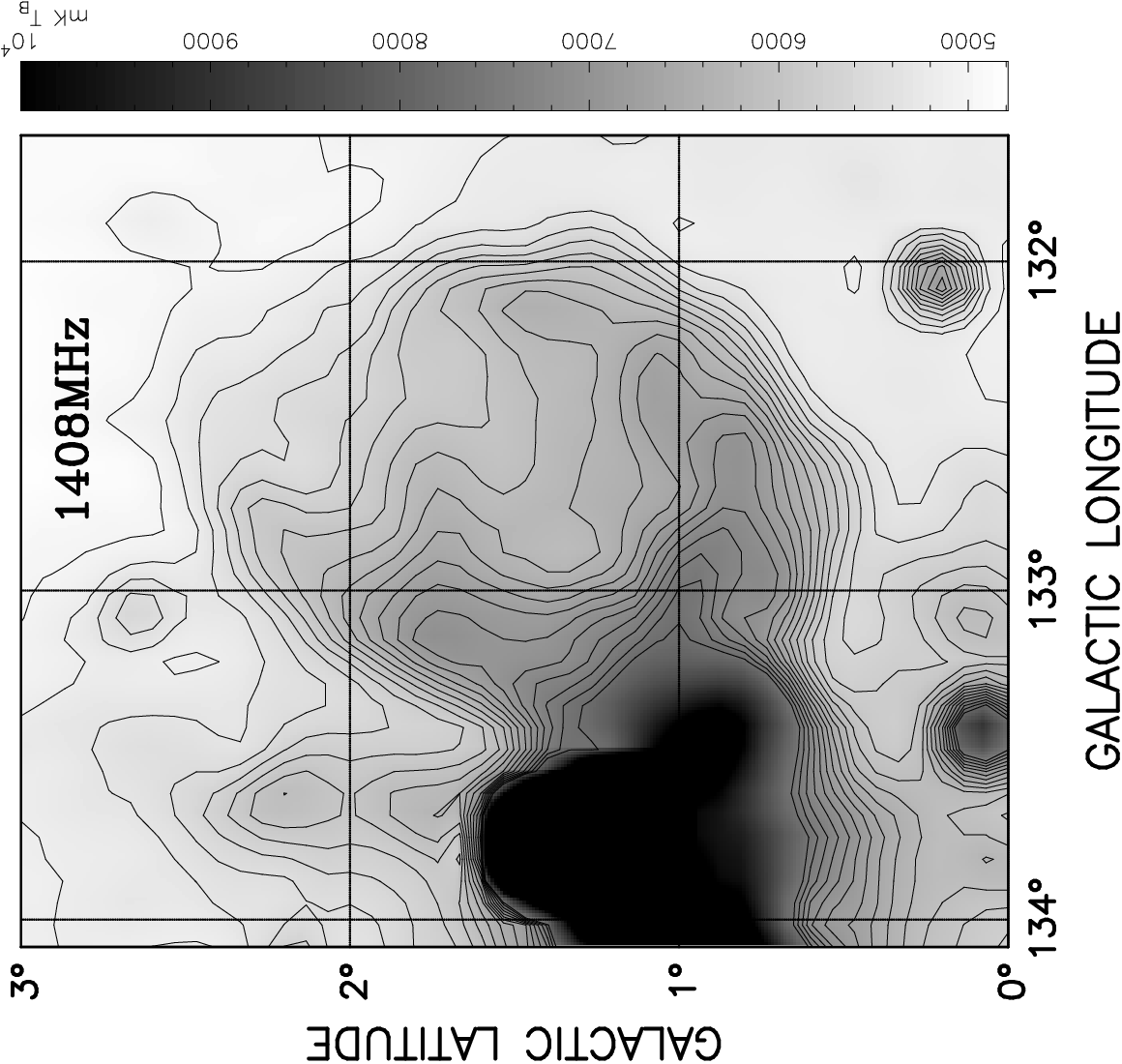}
\includegraphics[angle=270,width=0.30\textwidth]{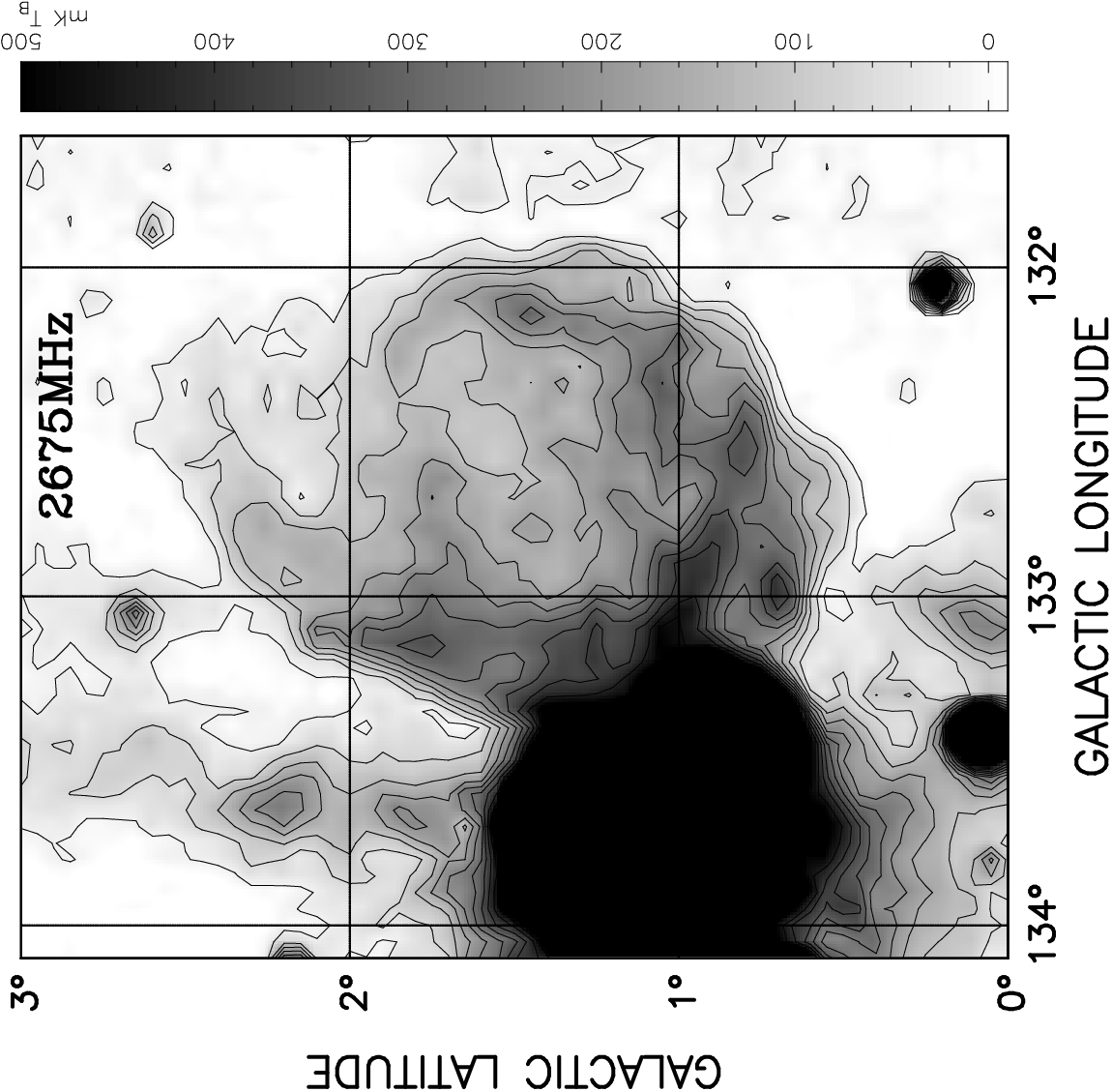}
\includegraphics[angle=270,width=0.30\textwidth]{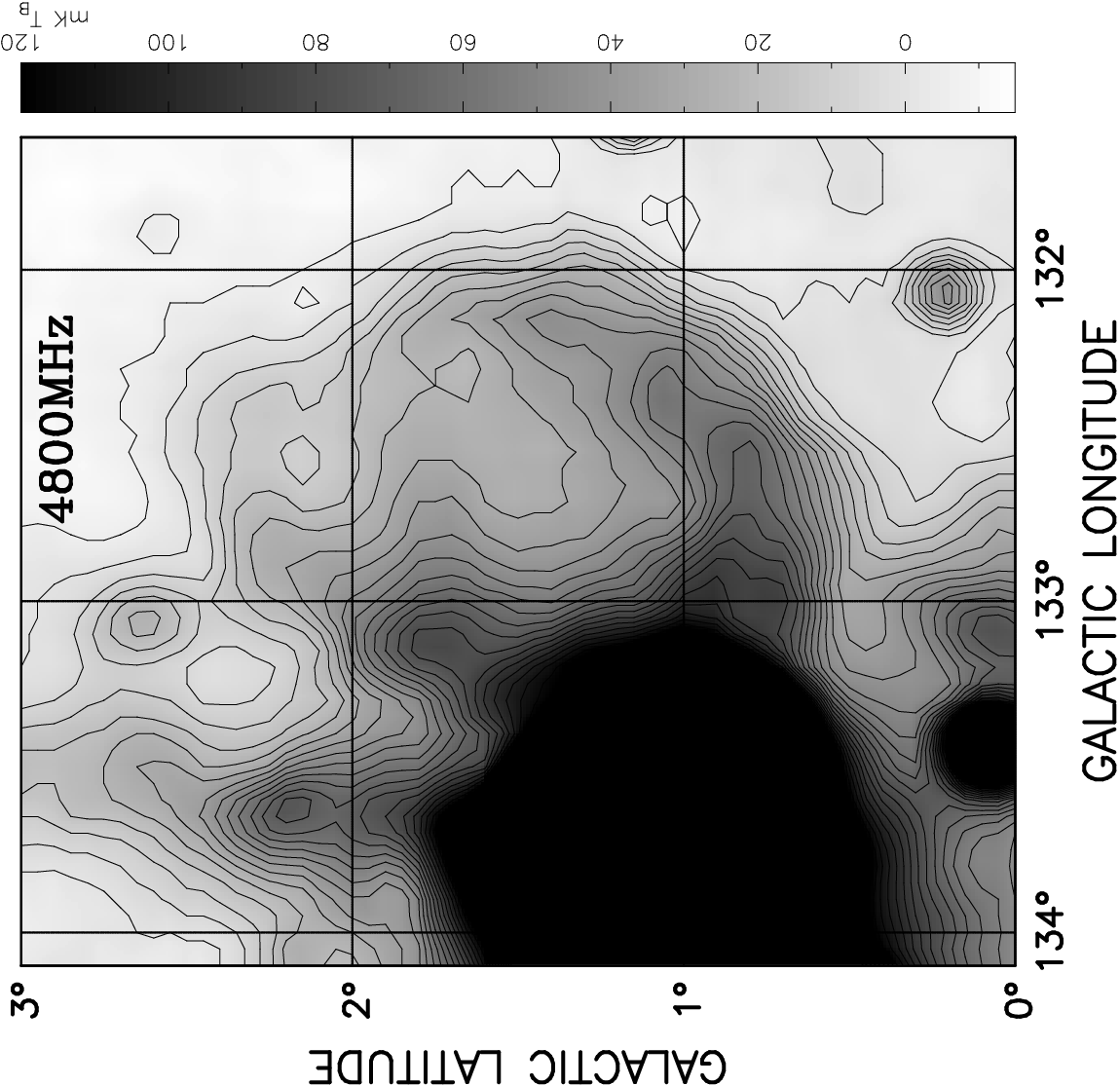} \\[2mm]
\includegraphics[angle=270,width=0.30\textwidth]{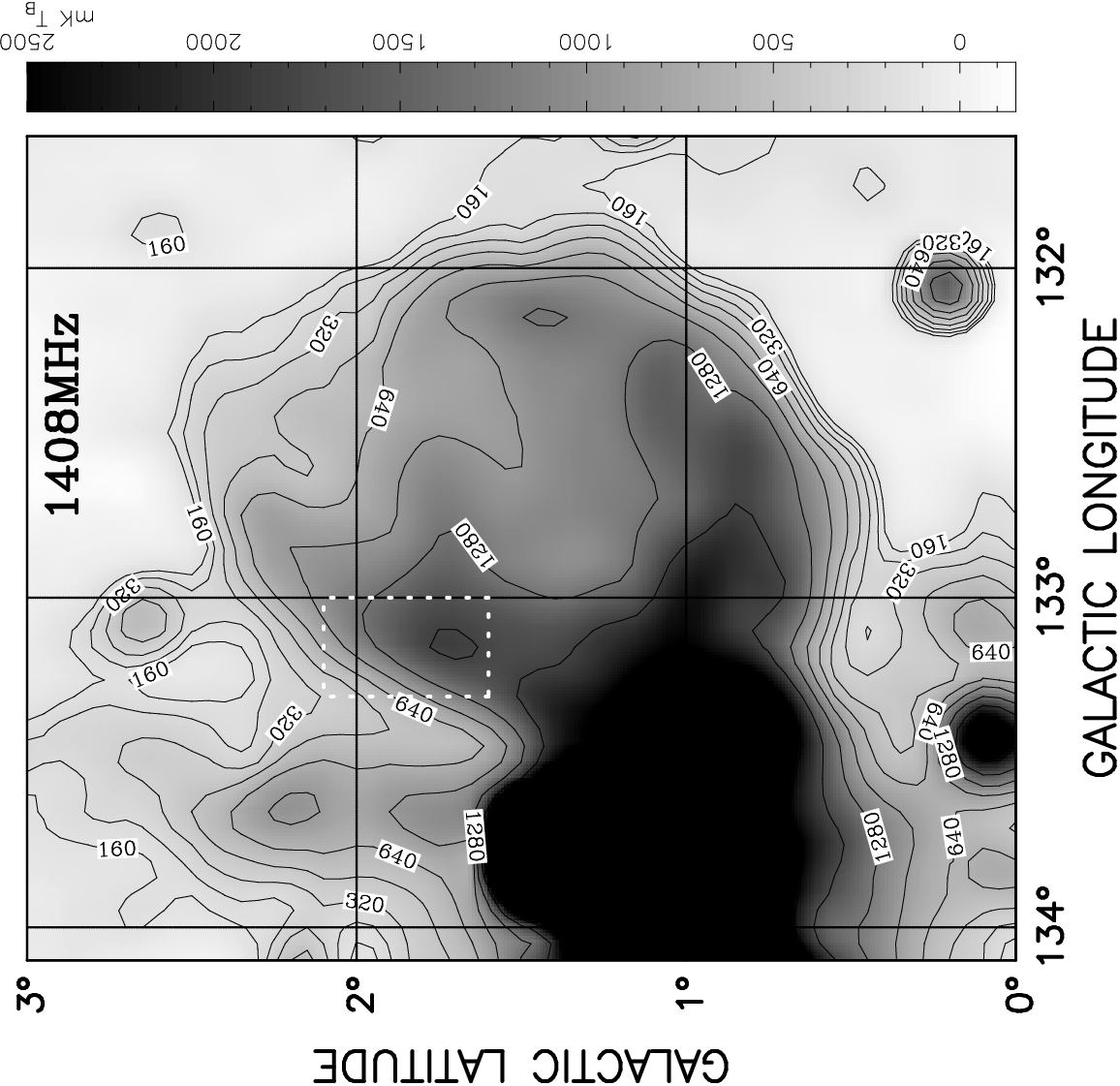}
\includegraphics[angle=270,width=0.30\textwidth]{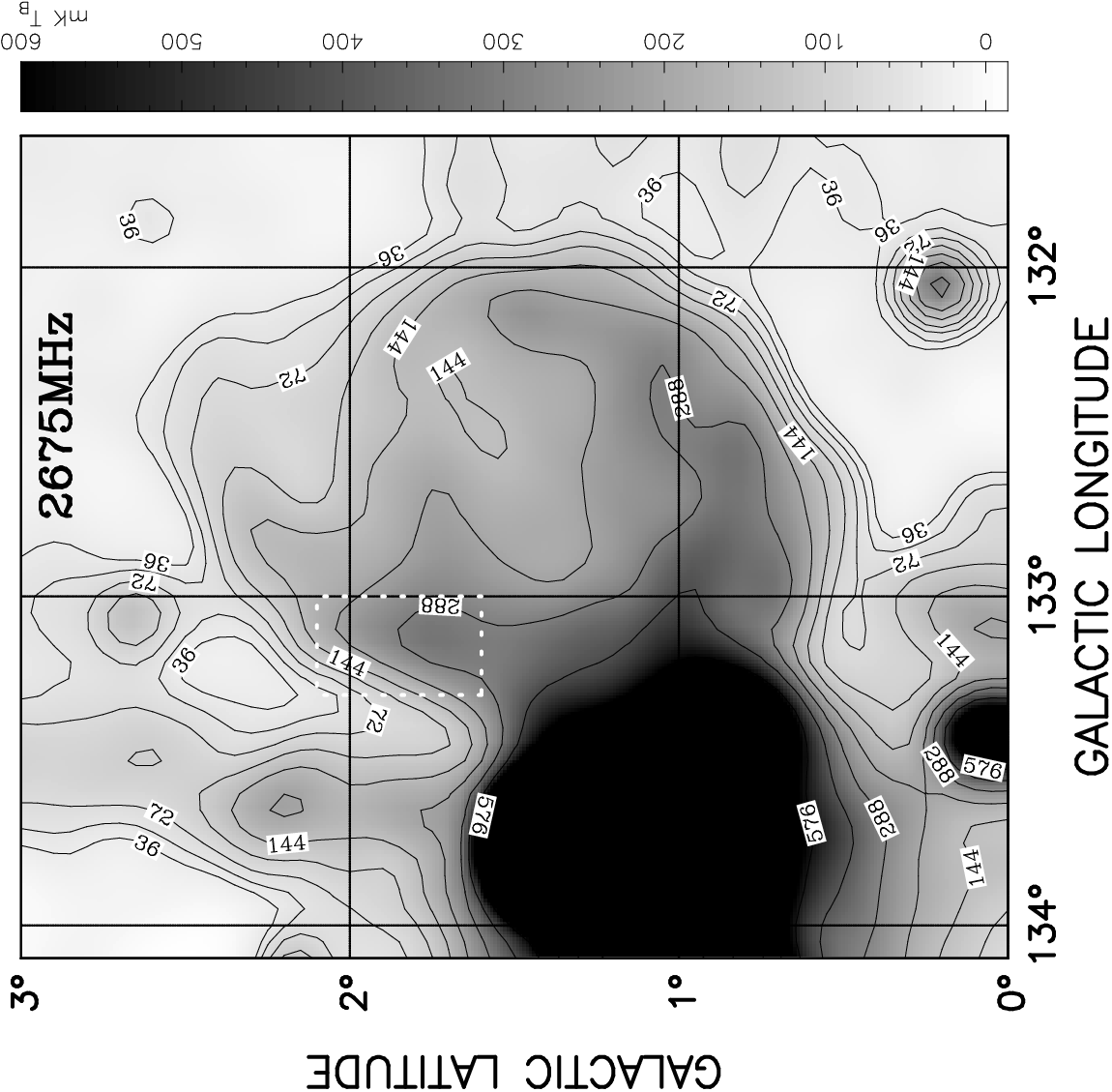}
\includegraphics[angle=270,width=0.30\textwidth]{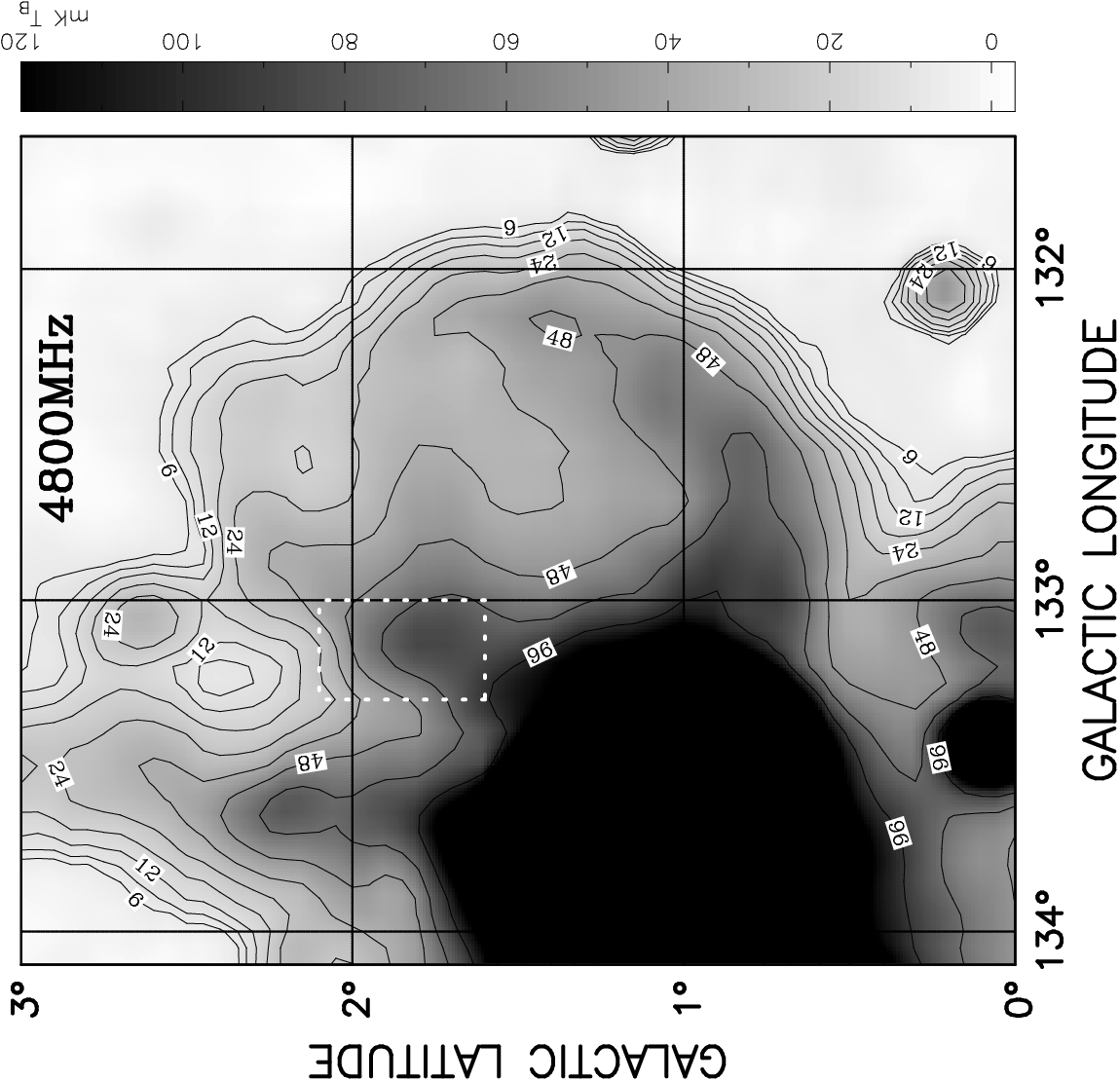}\\
\caption{HB3 maps at 1408~MHz, 2675~MHz, and 4800~MHz. The top row shows the
data after the subtraction of point-like sources in the area of HB3 and
$10\arcmin$ outside. The angular resolution at 1408 MHz, 2675 MHz, and 4800
MHz is $9\farcm4$, $4\farcm3$, and $9\farcm5$, respectively.  The bottom row
displays the same data convolved to a common angular resolution of
$10\arcmin$.}
\label{fig1}
\end{figure*}

HB3 is a large ($1\fdg5\times2\degr$) and evolved SNR, with an age of about
30000~yr \citep{ls06}. It is located at the edge of the HII regions W3/W4
and belongs to a complex of molecular clouds, star-formation and HII regions
(W3/W4/W5) \citep[e.g.,][]{dlp+96} in the Perseus arm at a distance of about
2.2~kpc \citep{rdlv91,xrzm06}. Although the SNR HB3 partly overlaps with the
adjacent HII regions, W3/W4, in the low-resolution radio maps, signs of a
real interaction, such as a deformed shell in the high-resolution maps or OH
masers in the joint region from the HB3 shock, have not been detected
\citep{kfg+98,kffu06}. Nevertheless, enhanced radio emission has been
observed at all radio bands in the southeastern regions of HB3 (towards the
lower-left in Fig.~\ref{fig1}). 

Previous estimates of the spectral index of HB3 were made from flux
densities determined at frequencies up to 3~GHz, mostly without point source
subtraction. The spectral index of HB3 was quoted to have values of 
$\alpha=-0.64$$\pm$$0.01$ \citep{fdwn95}, $\alpha=-0.56$$\pm$$0.03$ \citep{gre07},
$\alpha=-0.60$$\pm$$0.04$ \citep{lvdr87}, and $\alpha=-0.66$$\pm$$0.02$
\citep{kffu06}. Flux-density estimates of HB3 from low-resolution
observations are uncertain, especially at high frequencies, because of the
partial overlap with W3/W4 \citep[see][]{gre07}.

Recently, \citet{tl05} obtained images of HB3 at 408~MHz and 1420~MHz from
the Canadian Galactic Plane Survey \citep[CGPS: ][]{tgp+03} with compact
sources and diffuse background emission subtracted.  They derived a spectral
index for HB3 of $\alpha = -0.34$$\pm$$0.25$ between these two frequencies
\citep[][erratum]{tl05,tl06}. This result was confirmed by the TT-plot
method.  They also found a surprising result: {\it a spectral flattening of
the overall HB3 spectrum above 1~GHz}, derived from all published flux
density data from 38~MHz to 3900~MHz.  \citet{upl07} fitted the spectrum
using a two-component model of synchrotron emission and thermal
bremsstrahlung emission. They and concluded that the flattening is evidence
for the detection of thermal bremsstrahlung emission produced in the shell
of the SNR.  However, there are severe problems in the determination of flux
density data of HB3, as discussed by \citet{gre07}. The uncertainties in the
quoted flux density data, especially at high frequencies (1~GHz to 3~GHz),
are large, therefore the spectral flattening has been discredited by
\citet{gre07}. Radio observations at even higher frequencies are required to
clarify this extraordinary spectral behaviour.

The ongoing {\it Sino-German $\lambda$6~cm polarisation survey of the
Galactic plane} \citep{shr+07} has already covered the HB3 area. In this
paper, we analyse our $\lambda$6~cm (4800~MHz) data together with 1408~MHz
and 2675~MHz data from the Effelsberg Galactic plane surveys
\citep{rrf97,frrr90} to determine the high-frequency spectrum of HB3.
Furthermore, instead of using radio emission of the entire SNR, where its
eastern part obviously suffers from contamination or confusion by thermal
emission, we limit our study to the part being almost free of distortions.
The first detection of polarised emission of HB3 at $\lambda$6~cm will 
also be briefly discussed.

\section{Radio data of the HB3 region}

The 1408~MHz data are extracted from the Galactic plane 
survey\footnote {http://www.mpifr.de/survey.html} observed with the 
Effelsberg 100~m telescope \citep{rrf97}. The 2675 MHz data were 
observed in 1998 in the same way as the early 11cm survey \citet{frrr90}
but with a better receiver and higher sensitivity. The angular
resolutions are $9\farcm4$ and $4\farcm3$, respectively. The new 4800~MHz
data were observed with the Urumqi 25~m radio telescope with an angular
resolution of $9\farcm5$.  Observations were made by scans over $10\degr$ in
length in both Galactic longitude and latitude, which is the limiting size
for a reliable flux density determination of an object.  Technical details
of the 4800~MHz observations and data processing are described by
\citet{shr+07}.

For a more accurate determination of the radio emission of HB3 at these
three frequencies, we first identified and then subtracted unresolved point
sources above the local surrounding level of diffuse emission in the
region of HB3 and $10\arcmin$ outside. In total, 107 point sources have been
identified on the high-resolution CGPS maps at 408~MHz and 1420~MHz
\citep{tgp+03},  and their spectral indices were determined if visible
at both frequencies. Their fluxes at 2675~MHz and 4800~MHz were obtained by
extrapolation.  However, the fluxes of seven compact
sources at 2675~MHz have been taken from \citet{frrr90a} and
of eight point-like sources at 4800~MHz from the 87GB catalogue of
\citet{gc91}. The resulting maps are shown in the upper part of
Fig.~\ref{fig1}.

The large-scale Galactic background/foreground emission is obviously not
uniform in the maps. Clearly it is stronger in the eastern part than in the
western part, and stronger in the south than in the north. We subtracted a
``twisted'' base-level, so that the intensity level of the area surrounding
HB3 is close to zero, hence, removed the large-scale Galactic emission.  We
then convolved all maps to the same $10\arcmin$ angular resolution for the
spectral index analysis.

\begin{table}
\tabcolsep 1mm
\centering
\caption{Spectral indices of the HB3 regions}
\label{tab1}
\begin{tabular}{rccccccc}
\hline\hline
Freq. pairs&$\alpha_{\rm 133west}$&
$\beta_{\rm 133west}^{\rm TT}$&$\alpha_{\rm east.box}$&
$\beta_{\rm east.box}^{\rm TT}$\\
\hline
1408/2675&$-0.51\pm$0.16&$-2.64\pm$0.09&$-0.49\pm$0.25&$-2.69\pm$0.02\\
1408/4800&$-0.61\pm$0.06&$-2.61\pm$0.05&$-0.53\pm$0.09&$-2.54\pm$0.16\\
2675/4800&$-0.71\pm$0.18&$-2.57\pm$0.20&$-0.58\pm$0.28&$-2.32\pm$0.10\\
\hline
\end{tabular}
\end{table}

\begin{figure}[tbh]
\centering
\includegraphics[angle=270,width=0.4\textwidth]{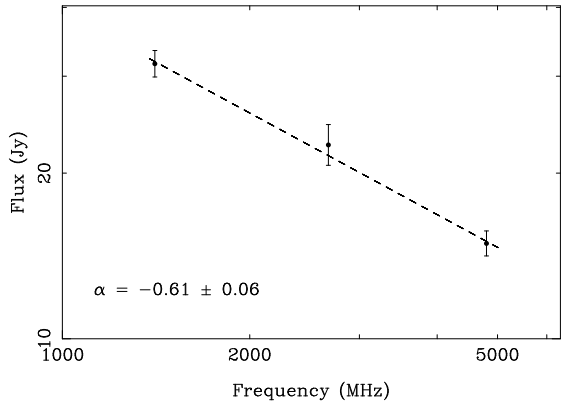}
\caption{Spectrum of the HB3 region west of $l=133\degr$.}
\label{fig2}
\end{figure}

\section{Spectral analysis of HB3}

The boundary between HB3 and the adjacent HII regions can be fairly well
defined in high-resolution low-frequency maps \citep{gre07}, but this is
difficult for our low-resolution maps. To avoid the contamination problem
that \citet{gre07} emphasised, the HB3 region west of $l=133\degr$ and in an
eastern box region (defined as $l=133\degr$ to $133\fdg3$, $b=1\fdg6$ to
$2\fdg1$) are analysed quantitatively.  No contamination from W3/W4 appears
in these regions for all three frequencies (see Fig.~\ref{fig1}).

\begin{figure*}[ht]
\centering
\includegraphics[angle=270,width=0.30\textwidth]{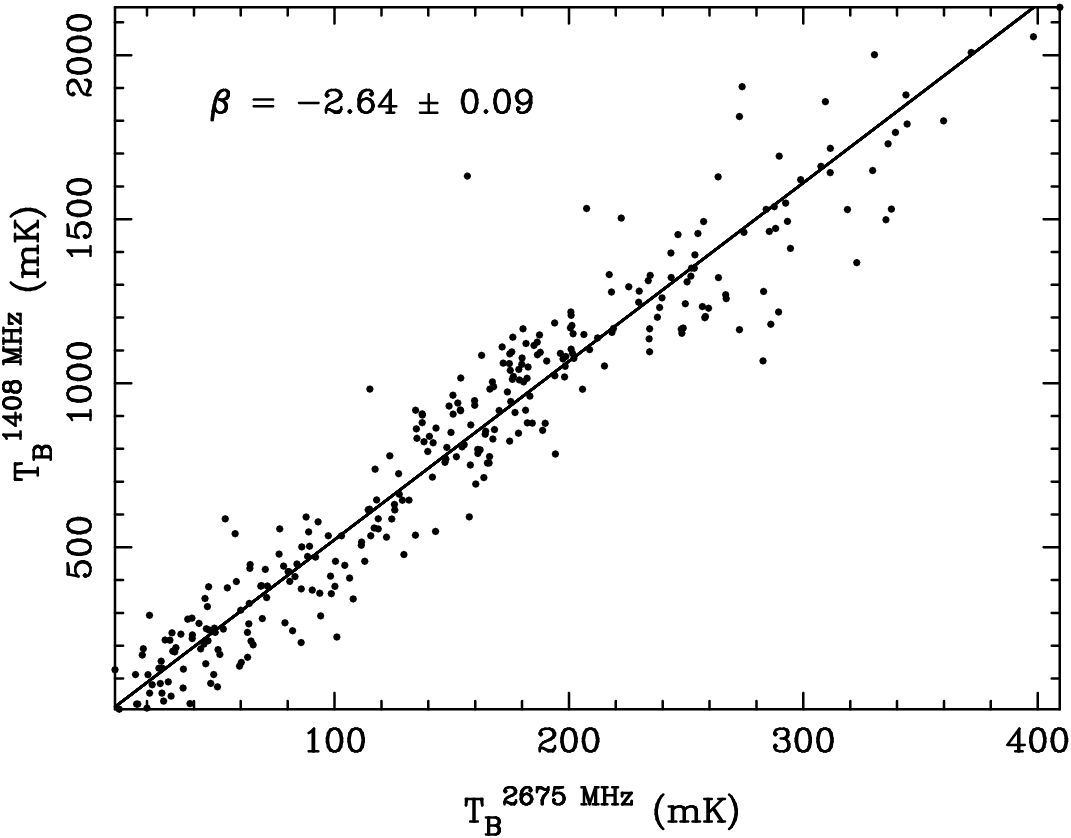}
\includegraphics[angle=270,width=0.30\textwidth]{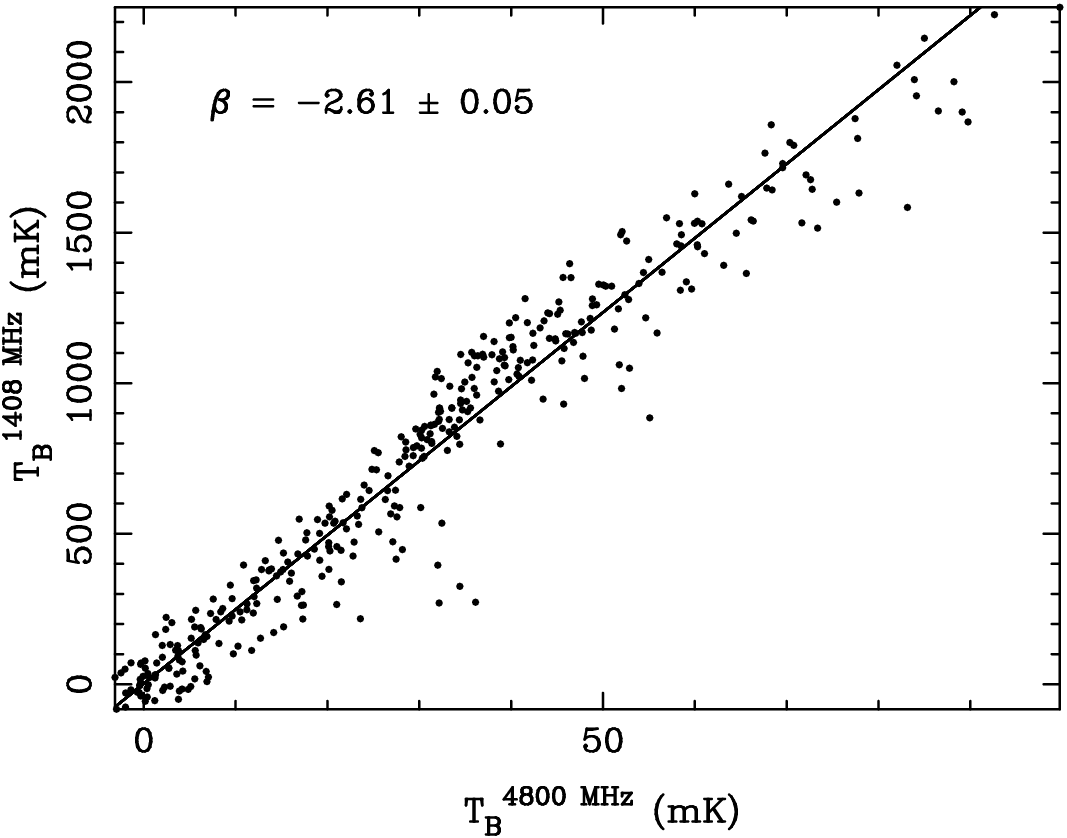}
\includegraphics[angle=270,width=0.30\textwidth]{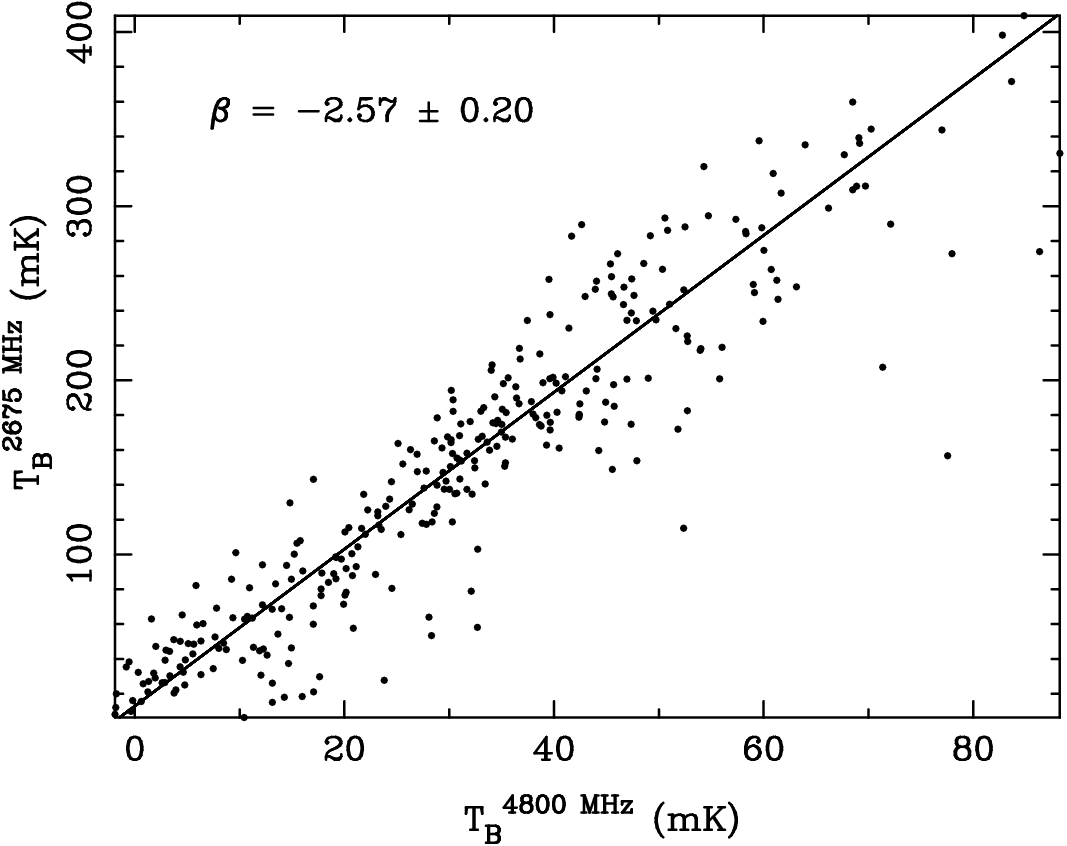}
\caption{TT-plots of the HB3 region west of $l=133\degr$.}
\label{fig3}
\end{figure*}


\begin{figure*}
\centering
\includegraphics[angle=270,width=0.30\textwidth]{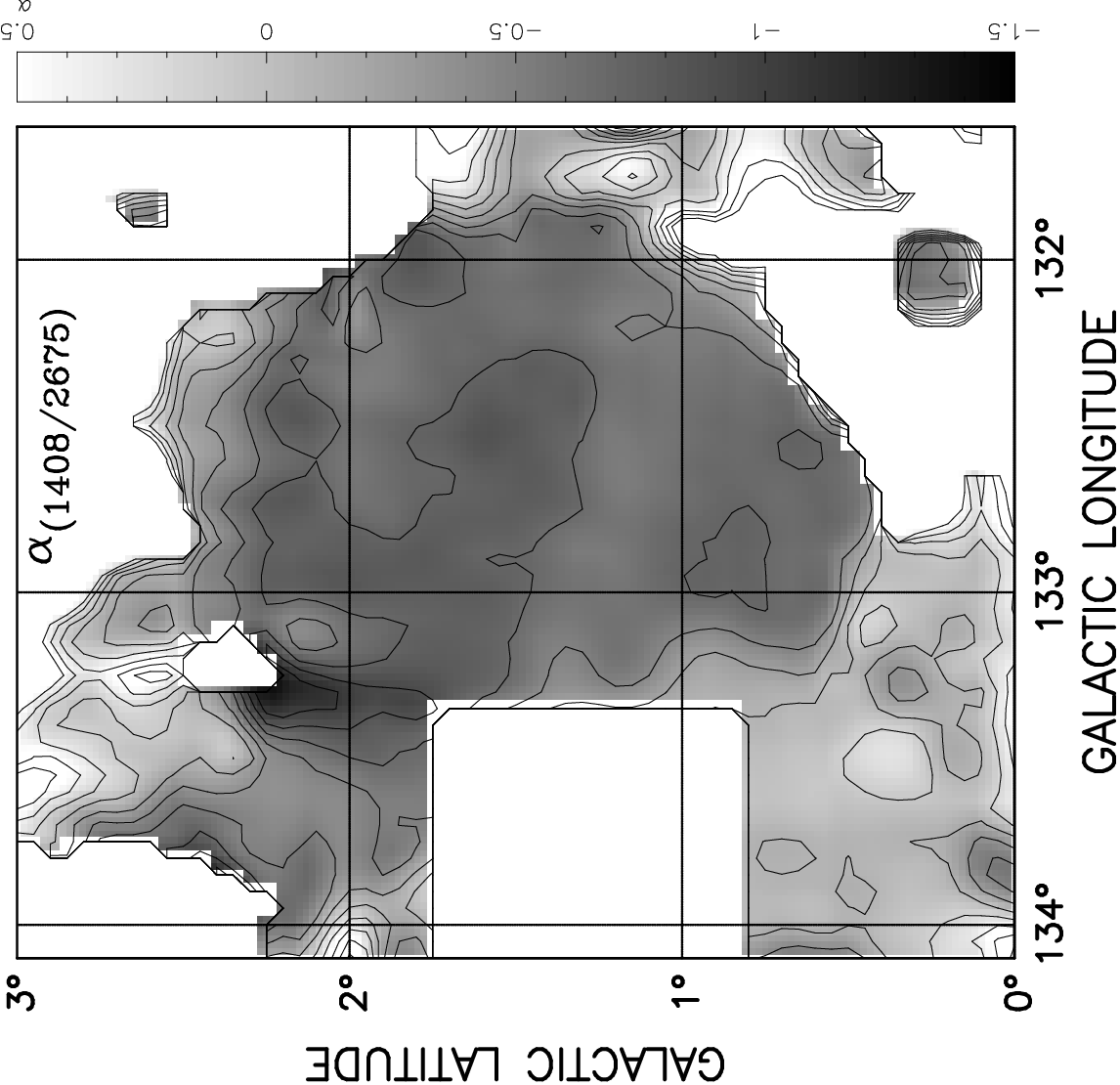}
\includegraphics[angle=270,width=0.30\textwidth]{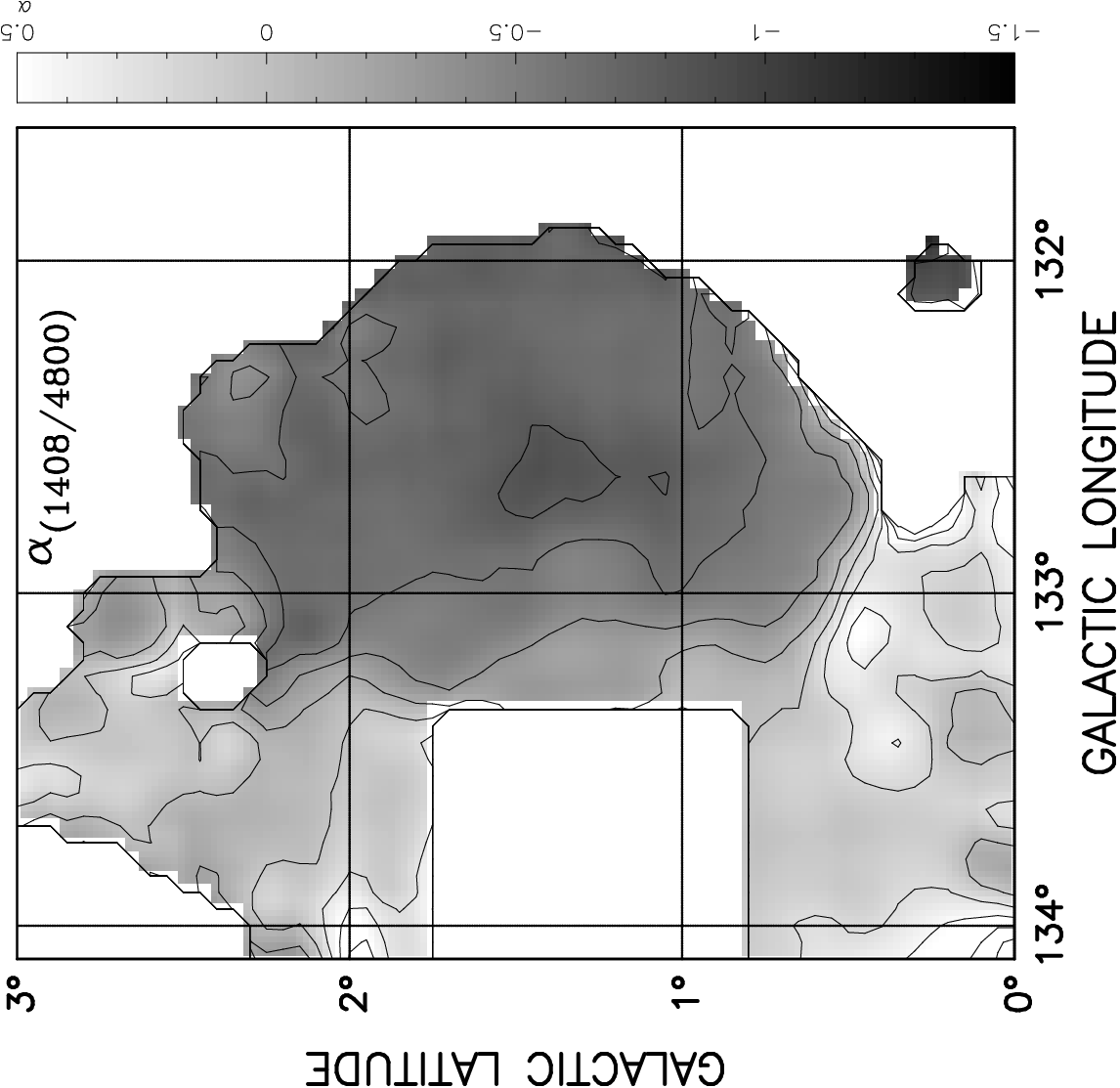}
\includegraphics[angle=270,width=0.30\textwidth]{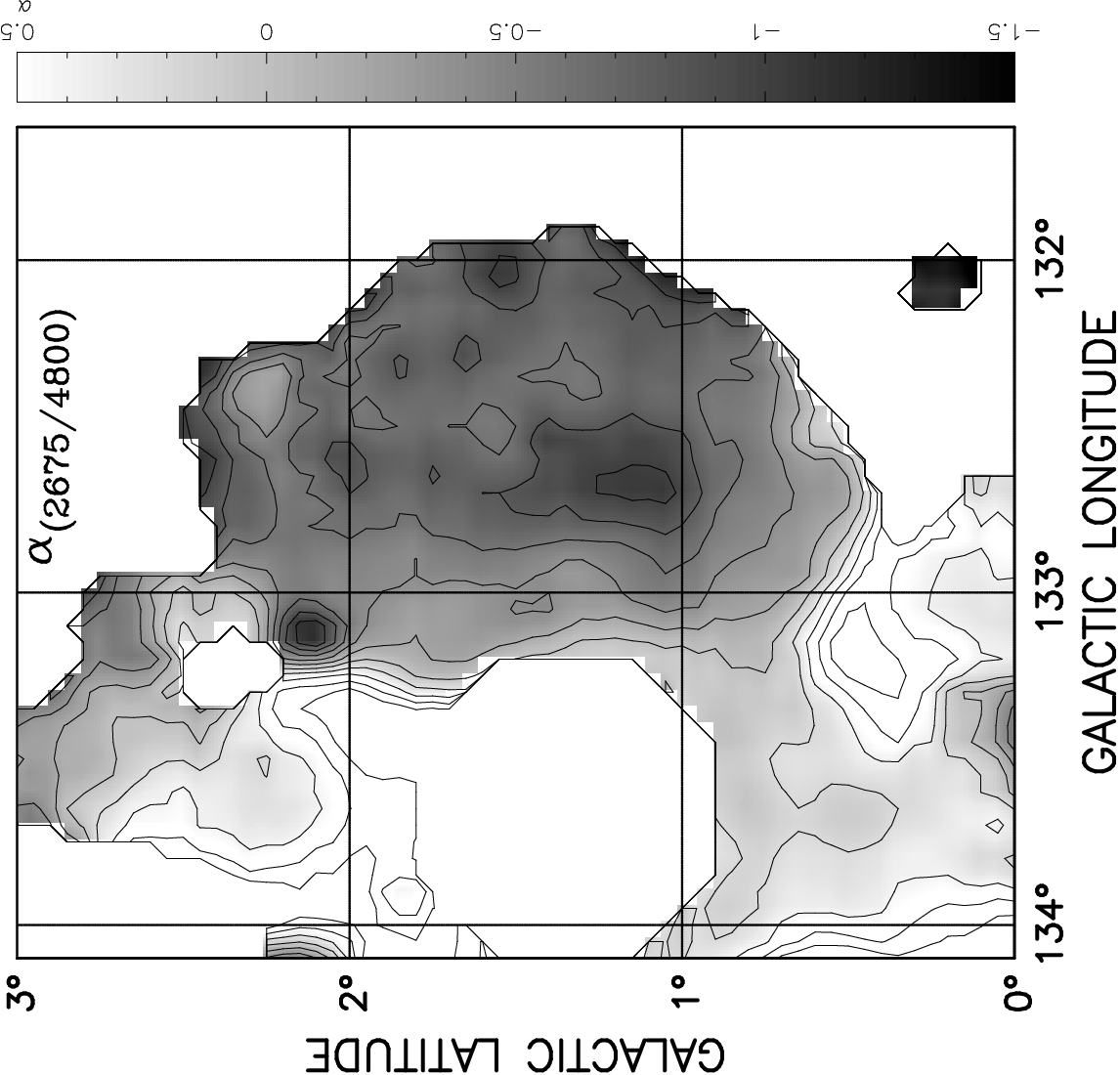}
\caption{Spectral index maps of HB3 calculated for three frequency pairs.}
\label{fig4}
\end{figure*}

\begin{figure*}
\centering
\includegraphics[angle=270,width=0.30\textwidth]{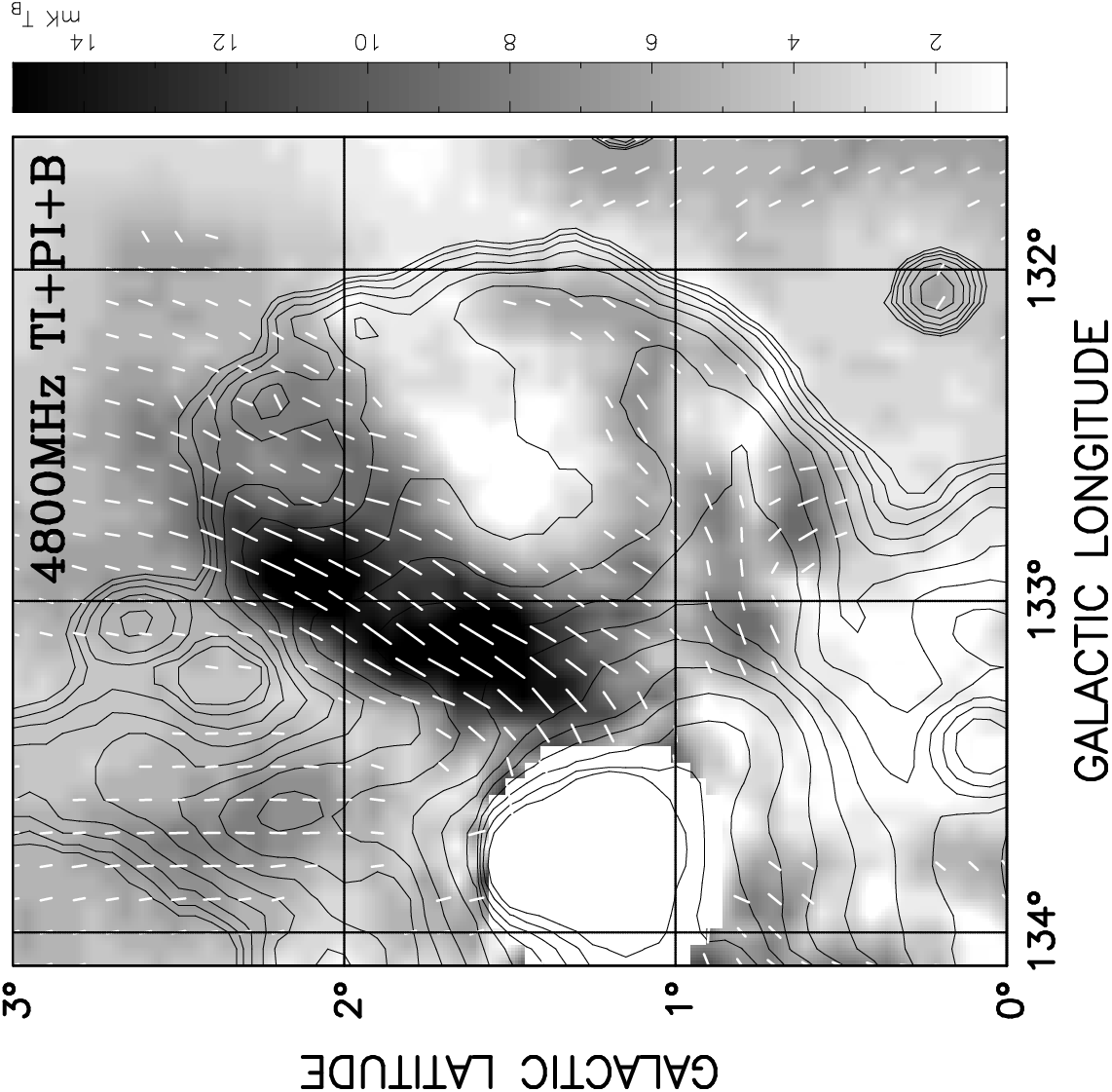}
\includegraphics[angle=270,width=0.30\textwidth]{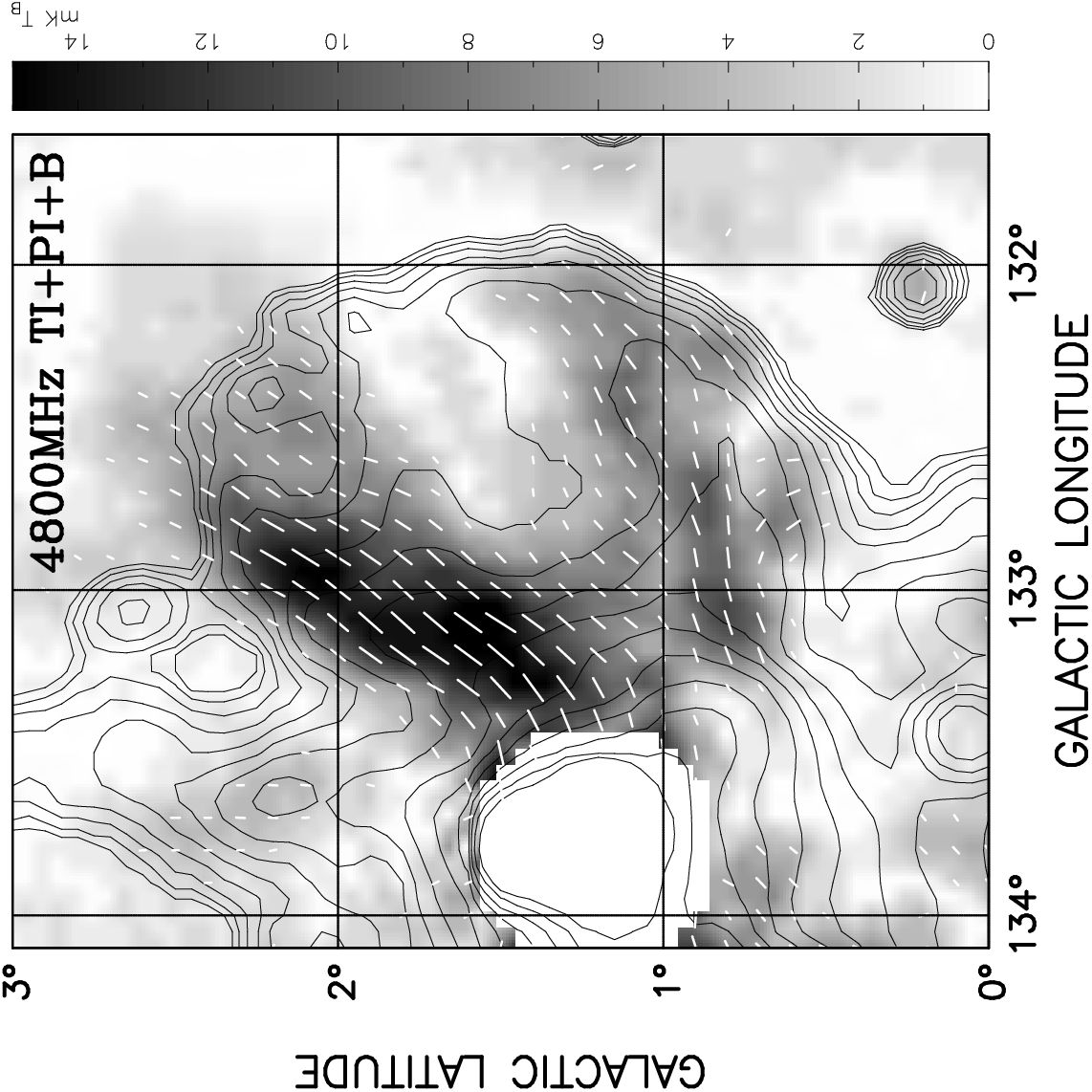}
\includegraphics[angle=270,width=0.30\textwidth]{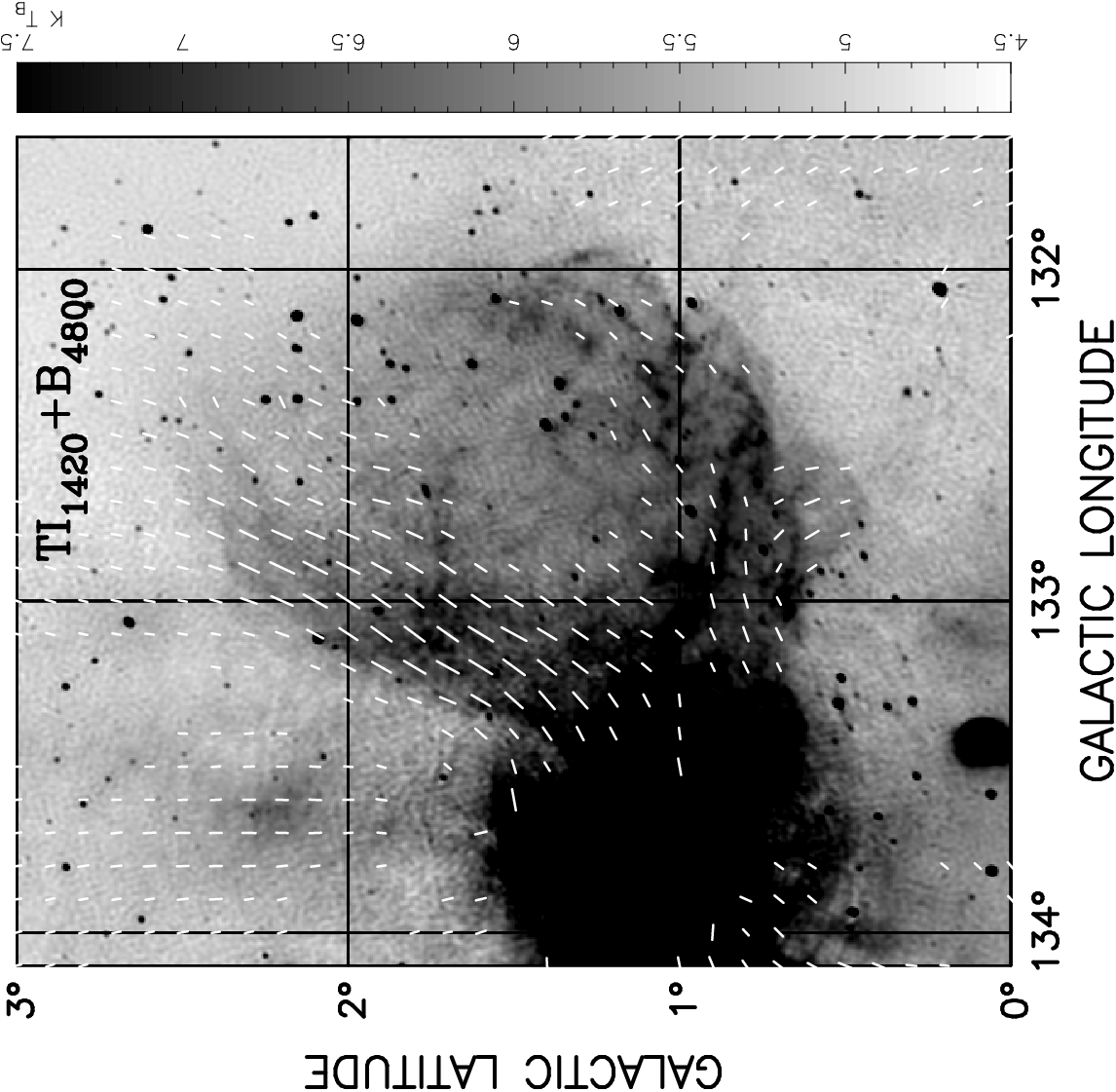}
\caption{Polarised emission from HB3 detected at 4800~MHz. {\it Left panel:}
Total intensity map at 4800~MHz is displayed by contours, and the polarised
intensity map is shown in grey-scale. The bars represent
($\vec{E}+90\degr$), roughly indicating the magnetic field orientation. The
bar length is proportional to the polarised intensity. The diffuse Galactic
emission is not subtracted. {\it Middle panel:} Same as the {\it left} but
diffuse Galactic polarised emission has been removed from the polarised
intensity map. {\it Right panel:} Polarisation angles ($\vec{E}+90\degr)
\simeq \vec{B}$ calculated from original maps at 4800~MHz are plotted over
the high-resolution 1420~MHz total intensity map (grey-scale) from the
Canadian Galactic Plane Survey \citep{tgp+03}.}
\label{fig5}
\end{figure*}

Fig.~\ref{fig2} shows that the integrated flux densities of the HB3 region
west of $l=133\degr$ decrease with frequency. 
The spectral index between 1408~MHz and 4800~MHz is $\alpha \sim -0.6$,
consistent with the value \citet{lvdr87} obtained from 408~MHz and 1420~MHz
maps, and that from the low-frequency integrated flux densities in
\citet{kffu06} and in Fig.3 of \citet{tl05}.

The HB3 spectral behaviour can be verified by the so called TT-plot method
(see Fig.~\ref{fig3}), which largely reduces the influence of unrelated
large-scale emission. Having convolved the maps to the same angular
resolution, the pixel values at every $5\arcmin$ sampling for a
selected area are plotted against each other.  The slope $\beta$ determined
by linear regression is the brightness temperature spectral index ($T
\propto \nu^{\,\beta}$).  The flux density spectral index, $\alpha$, is
related to $\beta$ as $\beta=\alpha-2$. The $\alpha$ and $\beta$ values in
Figs.~\ref{fig2} and \ref{fig3} are consistent (see Table~\ref{tab1}). The
emission in the eastern box region has a spectral behaviour similar to the
HB3 region west of $l=133\degr$.

The basic assumption for a reliable TT-plot method is that the emission at
each pixel of the extended object has the same spectral index between the
two frequencies, though the intensity varies from pixel to pixel. This is
the case for many SNRs, but is not always true. Deviations from the fitted
slope of a TT-plot may indicate either a possible spectral deviation for a
certain region or base-level fluctuations across the object. In
Fig.~\ref{fig4}, we show the resolved spectral index maps. The very central
part of HB3 seems to have a steeper spectrum, although the low intensity
there also implies a larger uncertainty in the spectral index.


\section{Polarised radio emission}

From our 4800~MHz observations, we clearly detected polarised emission from
HB3 for the first time (see Fig.~\ref{fig5}).  No polarised emission has
been detected at lower frequencies, so Faraday rotation of the polarisation
angles cannot be corrected. The polarised emission of HB3 at 4800~MHz is
strongest in the eastern shell with a percentage polarisation of up to
28\%. Some large-scale, polarised diffuse Galactic foreground emission is
also visible in the map.

Pulsar rotation measures in the direction of the Perseus arm are a few tens
of rad~m$^{-2}$ at maximum \citep{hmq99}. This results in a polarisation
angle rotation of less than $20\degr$ at 4800~MHz.  A $90\degr$ rotation of
our observed $\vec{E}$ vectors roughly indicates the magnetic field
orientation. Magnetic field lines seem to follow the shell, including the
weak polarised emission in the western limb, which is a typical
characteristic of an evolved SNR. A very exceptional area showing a radial
magnetic field is located in the extreme southern part of the rim.  It
resembles a circular half shell with a diameter of $20\arcmin$, centred at
$(l,b)=(132\fdg75,0\fdg55)$ (see {\it the right panel} of Fig.~\ref{fig5}, and
also the H$\alpha$ image, Fig.~8, of \citealt{fdwn95}). Based on the
distinct magnetic morphology, we suspect this may be another young SNR,
rather than a limb of HB3. However, higher-resolution polarisation
observations are needed to clarify the nature of this feature.

In the central region, where complete depolarisation occurs in our 4800~MHz
map, ring-like X-ray emission has been detected by the Einstein satellite
\citep[for a comparison with the radio emission see ][]{lvdr87,vlg+84} and
more recent ROSAT observations \citep[see Fig.11 in ][]{ls06}.

\section{Discussion and Conclusions}

We studied the spectrum of HB3 using maps at 1408~MHz, 2675~MHz, and 4800~MHz
for the regions of HB3 where no contamination from W3/W4 is visible.  We
obtained a spectral index for HB3 between 1408~MHz and 4800~MHz of $\alpha =
-0.61$$\pm$$0.06$, which is consistent with values obtained by
\citet{fdwn95}, \citet{lvdr87}, \citet{kffu06}. The TT-plot method was used
to verify the spectrum. The new 4800~MHz data directly show that the
spectrum of HB3 is constant with $\alpha \sim -0.6$ from 1~GHz to 5~GHz.
There is no evidence for a spectral flattening to higher frequencies
discussed earlier by \citet{tl05}. This conclusion is in entire agreement
with \citet{gre07}. Therefore, the thermal bremsstrahlung emission
discussed by \citet{upl07} does not exist in the HB3 areas not contaminated
by thermal emission related to W3/W4.

Polarised radio emission is clearly detected, which indicates that the
magnetic field is aligned with the shell of HB3, except for the southern
spherical limb, which we suspect to be an independent feature.

\begin{acknowledgements}
The $\lambda$6\ cm data were obtained with the receiver system from the
MPIfR mounted at the Nanshan 25~m telescope at the Urumqi Observatory of
NAOC. The authors are supported by the National Natural Science Foundation
(NNSF) of China (10521001 and 10773016), the National Key Basic Research
Science Foundation of China (2007CB815403), and the Partner group of the
MPIfR at NAOC in the frame of the exchange program between MPG and CAS for
many bilateral visits. We thank Dr. James Anderson for carefully reading the
manuscript.
\end{acknowledgements}



\begin{thebibliography}{22}
\expandafter\ifx\csname natexlab\endcsname\relax\def\natexlab#1{#1}\fi

\bibitem[{{Digel} {et~al.}(1996){Digel}, {Lyder}, {Philbrick}, {Puche}, \&
  {Thaddeus}}]{dlp+96}
{Digel}, S.~W., {Lyder}, D.~A., {Philbrick}, A.~J., {Puche}, D., \& {Thaddeus},
  P. 1996, ApJ, 458, 561

\bibitem[{{Fesen} {et~al.}(1995){Fesen}, {Downes}, {Wallace}, \&
  {Normandeau}}]{fdwn95}
{Fesen}, R.~A., {Downes}, R.~A., {Wallace}, D., \& {Normandeau}, M. 1995, AJ,
  110, 2876

\bibitem[{{F\"urst} \& {Reich}(1986)}]{fr86}
{F\"urst}, E., \& {Reich}, W. 1986, A\&A, 163, 185

\bibitem[{{F\"urst} {et~al.}(1990{\natexlab{a}}){F\"urst}, {Reich}, {Reich}, \&
  {Reif}}]{frrr90}
{F\"urst}, E., {Reich}, W., {Reich}, P., \& {Reif}, K. 1990{\natexlab{a}},
  A\&AS, 85, 691

\bibitem[{{F\"urst} {et~al.}(1990{\natexlab{b}}){F\"urst}, {Reich}, {Reich}, \&
  {Reif}}]{frrr90a}
{F\"urst}, E., {Reich}, W., {Reich}, P., \& {Reif}, K. 1990{\natexlab{b}},
  A\&AS, 85, 805

\bibitem[{{Green}(2007)}]{gre07}
{Green}, D.~A. 2007, Bulletin of the Astronomical Society of India, 35, 77

\bibitem[{{Gregory} \& {Condon}(1991)}]{gc91}
{Gregory}, P.~C. \& {Condon}, J.~J. 1991, ApJS, 75, 1011

\bibitem[{{Han} {et~al.}(1999){Han}, {Manchester}, \& {Qiao}}]{hmq99}
{Han}, J.~L., {Manchester}, R.~N., \& {Qiao}, G.~J. 1999, \mnras, 306, 371

\bibitem[{{Koralesky} {et~al.}(1998){Koralesky}, {Frail}, {Goss}, {Claussen},
  \& {Green}}]{kfg+98}
{Koralesky}, B., {Frail}, D.~A., {Goss}, W.~M., {Claussen}, M.~J., \& {Green},
  A.~J. 1998, AJ, 116, 1323

\bibitem[{{Kothes} {et~al.}(2006){Kothes}, {Fedotov}, {Foster}, \&
  {Uyan{\i}ker}}]{kffu06}
{Kothes}, R., {Fedotov}, K., {Foster}, T.~J., \& {Uyan{\i}ker}, B. 2006, A\&A,
  457, 1081

\bibitem[{{Landecker} {et~al.}(1987){Landecker}, {Vaneldik}, {Dewdney}, \&
  {Routledge}}]{lvdr87}
{Landecker}, T.~L., {Vaneldik}, J.~F., {Dewdney}, P.~E., \& {Routledge}, D.
  1987, AJ, 94, 111

\bibitem[{{Lazendic} \& {Slane}(2006)}]{ls06}
{Lazendic}, J.~S., \& {Slane}, P.~O. 2006, ApJ, 647, 350

\bibitem[{{Reich} {et~al.}(1997){Reich}, {Reich}, \& {F\"urst}}]{rrf97}
{Reich}, P., {Reich}, W., \& {F\"urst}, E. 1997, A\&AS, 126, 413

\bibitem[{{Routledge} {et~al.}(1991){Routledge}, {Dewdney}, {Landecker}, \&
  {Vaneldik}}]{rdlv91}
{Routledge}, D., {Dewdney}, P.~E., {Landecker}, T.~L., \& {Vaneldik}, J.~F.
  1991, A\&A, 247, 529

\bibitem[{{Sun} {et~al.}(2007){Sun}, {Han}, {Reich}, {Reich}, {Shi},
  {Wielebinski}, \& {F{\"u}rst}}]{shr+07}
{Sun}, X.~H., {Han}, J.~L., {Reich}, W., {et~al.} 2007, A\&A, 463, 993

\bibitem[{{Taylor} {et~al.}(2003){Taylor}, {Gibson}, {Peracaula}, {Martin},
  {Landecker}, {Brunt}, {Dewdney}, {Dougherty}, {Gray}, {Higgs}, {Kerton},
  {Knee}, {Kothes}, {Purton}, {Uyaniker}, {Wallace}, {Willis}, \&
  {Durand}}]{tgp+03}
{Taylor}, A.~R., {Gibson}, S.~J., {Peracaula}, M., {et~al.} 2003, AJ, 125, 3145

\bibitem[{{Tian} \& {Leahy}(2005)}]{tl05}
{Tian}, W.~W., \& {Leahy}, D. 2005, A\&A, 436, 187

\bibitem[{{Tian} \& {Leahy}(2006)}]{tl06}
{Tian}, W.~W., \& {Leahy}, D.~A. 2006, erratum, \aap, 451, 991

\bibitem[{{Uro{\v s}evi{\'c}} {et~al.}(2007){Uro{\v s}evi{\'c}}, {Pannuti}, \&
  {Leahy}}]{upl07}
{Uro{\v s}evi{\'c}}, D., {Pannuti}, T.~G., \& {Leahy}, D. 2007, ApJ, 655, L41

\bibitem[{{Venkatesan} {et~al.}(1984){Venkatesan}, {Leahy}, {Galas}, {Naranan},
  \& {Long}}]{vlg+84}
{Venkatesan}, D., {Leahy}, D.~A., {Galas}, C.~M.~F., {Naranan}, S., \& {Long},
  K. 1984, \mnras, 208, 25P

\bibitem[{{Xiao} {et~al.}(2008){Xiao}, {F{\"u}rst}, {Reich}, \& {Han}}]{xfrh08}
{Xiao}, L., {F{\"u}rst}, E., {Reich}, W., \& {Han}, J.~L. 2008, A\&A, 482, 783

\bibitem[{{Xu} {et~al.}(2006){Xu}, {Reid}, {Zheng}, \& {Menten}}]{xrzm06}
{Xu}, Y., {Reid}, M.~J., {Zheng}, X.~W., \& {Menten}, K.~M. 2006, Science, 311,
  54

\end{thebibliography}

\end{document}